\documentclass[sigconf,screen]{acmart}

\AtBeginDocument{%
  \providecommand\BibTeX{{%
    \normalfont B\kern-0.5em{\scshape i\kern-0.25em b}\kern-0.8em\TeX}}}

\usepackage{longfbox}
\usepackage{graphicx}
\usepackage{amsmath}
\usepackage[american]{babel}
\usepackage{microtype}
\usepackage{subcaption}
\usepackage[ruled,,linesnumbered ]{algorithm2e}
\usepackage[most]{tcolorbox}
\usepackage{balance} 
\usepackage{enumitem} 
\usepackage{color}
\usepackage{url}
\usepackage{arydshln}
\usepackage{bm}
\usepackage{multirow}
\usepackage{tikz}

\DeclareMathOperator*{\argmax}{\arg\max}
\newcommand{\doublecircled}[1]{%
  \tikz[baseline=(char.base)]{
    \node[shape=circle,draw,inner sep=0.3pt] (char) {#1};}%
}

\usepackage{tcolorbox}
\tcbuselibrary{skins, breakable, theorems}
\usepackage{colortbl}
\usepackage{geometry}

\newcommand{\system}{\textsc{CFExplainer \xspace}}
\newcommand{\systemnospace}{\textsc{CFExplainer}}

\setcopyright{acmlicensed}
\acmDOI{10.1145/3650212.3652136}
\acmYear{2024}
\copyrightyear{2024}
\acmSubmissionID{issta24main-p884-p}
\acmISBN{979-8-4007-0612-7/24/09}
\acmConference[ISSTA '24]{Proceedings of the 33rd ACM SIGSOFT International Symposium on Software Testing and Analysis}{September 16--20, 2024}{Vienna, Austria}
\acmBooktitle{Proceedings of the 33rd ACM SIGSOFT International Symposium on Software Testing and Analysis (ISSTA '24), September 16--20, 2024, Vienna, Austria}
\received{16-DEC-2023}
\received[accepted]{2024-03-02}

\author{Zhaoyang Chu}
\authornote{Also with National Engineering Research Center for Big Data Technology and System, Services Computing Technology and System Lab, Cluster and Grid Computing Lab, Huazhong University of Science and Technology, Wuhan, 430074, China.}
\affiliation{%
 \institution{School of Computer Science and Technology, Huazhong University of Science and Technology, China}
 \streetaddress{Luoyu Rd 1037}
 \country{}
 }
\email{chuzhaoyang@hust.edu.cn}

\author{Yao Wan}
\authornotemark[1]
\authornote{Yao Wan is the corresponding author.}
\affiliation{%
 \institution{School of Computer Science and Technology, Huazhong University of Science and Technology, China}
 \country{}
 }
\email{wanyao@hust.edu.cn}

\author{Qian Li}
\affiliation{%
\institution{School of Electrical Engineering, Computing and Mathematical Sciences, Curtin University, Australia}
\country{}
}
\email{qli@curtin.edu.au}

\author{Yang Wu}
\authornotemark[1]
\affiliation{%
 \institution{School of Computer Science and Technology, Huazhong University of Science and Technology, China}
 \country{}
 }
\email{wuyang_emily@hust.edu.cn}

\author{Hongyu Zhang}
\affiliation{%
\institution{School of Big Data and Software Engineering, Chongqing University, China}
\country{}
}
\email{hyzhang@cqu.edu.cn}

\author{Yulei Sui}
\affiliation{%
\institution{School of Computer Science and Engineering, University of New South Wales, Australia} %
\country{}
}
\email{y.sui@unsw.edu.au}

\author{Guandong Xu}
\affiliation{%
\institution{School of Computer Science, University of Technology Sydney, Australia} 
\country{}
}
\email{guandong.xu@uts.edu.au}

\author{Hai Jin}
\authornotemark[1]
\affiliation{%
 \institution{School of Computer Science and Technology, Huazhong University of Science and Technology, China}
 \country{}
 }
\email{hjin@hust.edu.cn}

\begin{document}

\title[Graph Neural Networks for Vulnerability Detection: A Counterfactual Explanation]{Graph Neural Networks for Vulnerability Detection: \\
A Counterfactual Explanation}

\begin{abstract}
Vulnerability detection is crucial for ensuring the security and reliability of software systems. 
Recently, \textit{Graph Neural Networks} (GNNs) have emerged as a prominent code embedding approach for vulnerability detection, owing to their ability to capture the underlying semantic structure of source code. 
However, GNNs face significant challenges in explainability due to their inherently black-box nature.
To this end, several \textit{factual reasoning}-based explainers have been proposed. 
These explainers provide explanations for the predictions made by GNNs by analyzing the key features that contribute to the outcomes.
We argue that these factual reasoning-based explanations cannot answer critical \emph{what-if} questions: ``\textit{What would happen to the GNN's decision if we were to alter the code graph into alternative structures?}'' 
Inspired by advancements of \textit{counterfactual reasoning} in artificial intelligence, we propose \systemnospace, a novel counterfactual explainer for GNN-based vulnerability detection. 
Unlike factual reasoning-based explainers, \system seeks the minimal perturbation to the input code graph that leads to a change in the prediction, thereby addressing the \emph{what-if} questions for vulnerability detection.
We term this perturbation a counterfactual explanation, which can pinpoint the root causes of the detected vulnerability and furnish valuable insights for developers to undertake appropriate actions for fixing the vulnerability.
Extensive experiments on four GNN-based vulnerability detection models demonstrate the effectiveness of \system over existing state-of-the-art factual reasoning-based explainers.
\end{abstract}



\maketitle

\section{Introduction}
\label{sec_intro}

Software vulnerabilities, which expose weaknesses in a program, present a significant risk to data integrity, user privacy, and overall cybersecurity~\cite{Li2018vuldeepecker, Li2021ivdetect, Zhou2019devign}.
As of now, the \textit{Common Vulnerabilities and Exposures} (CVE)~\cite{Gu2022hierarchical} has reported tens of thousands of software vulnerabilities annually.
Thus, vulnerability detection, which aims to automatically identify potentially vulnerable code, plays a pivotal role in ensuring the security and reliability of software. 

\begin{figure*}[!t]
    \centering
    \includegraphics[width=0.96\linewidth]{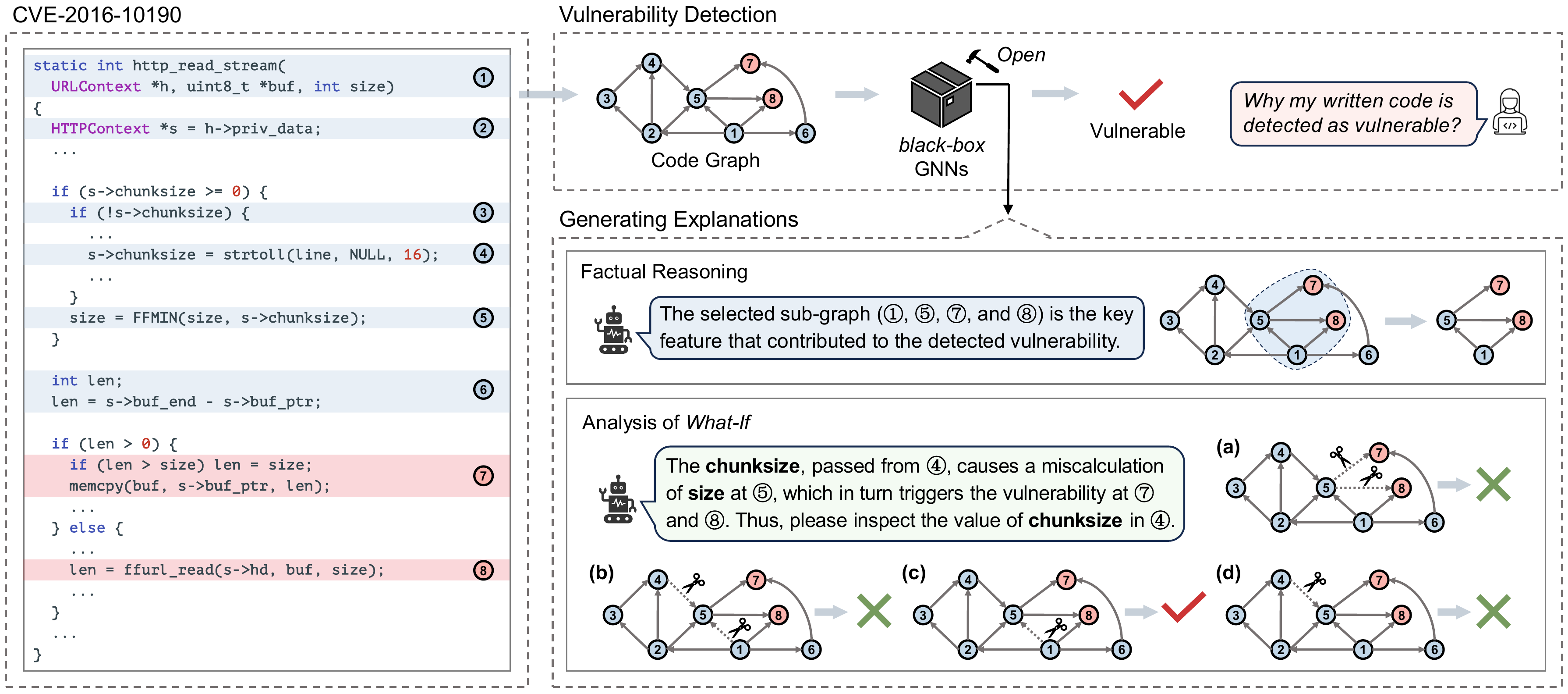}
    \vspace{-0.3em}
    \caption{Illustration of factual reasoning-based explanation (right middle) and \emph{what-if} analysis (right bottom).}
    \label{fig_illustration}
    \vspace{-0.5em}
\end{figure*}

Existing efforts on vulnerability detection primarily fall within two main categories: static analysis-based approaches~\cite{Gao2018cobot, Sui2016svf, Viega2000its4} and deep learning-based approaches~\cite{Hin2022linevd, Li2018vuldeepecker, Li2022vuldeelocator, Zhou2019devign}.
Traditional static analysis-based approaches (e.g., SVF~\cite{Sui2016svf} and Infer~\cite{infer}) rely on human experts to manually define specific rules for detecting vulnerabilities.
Recently, deep learning-based approaches, exemplified by pioneering works such as VulDeePecker~\cite{Li2018vuldeepecker} and Devign~\cite{Zhou2019devign}, have made remarkable strides, largely attributed to their capacity to learn comprehensive code representations, thereby enhancing the detection capabilities across diverse vulnerabilities.
Among these approaches, \textit{Graph Neural Networks} (GNNs)~\cite{Chakraborty2022reveal, Cheng2021deepwukong, Hin2022linevd, Li2021ivdetect, Zhou2019devign} have recently attracted substantial attention, owing to their capacity to capture intricate structural information of code, e.g., syntax trees, control flows, and data flows.

Despite the significant progress made by GNNs in vulnerability detection, existing detection systems suffer from the \textit{explainability} issues due to the black-box and complicated nature of deep neural networks.
Given a predicted result, developers are often confused by the following question: ``\textit{Why my code is detected as vulnerable?}''
From our investigation, existing studies~\cite{Hu2023vul_interpreters, Ganz2021vul_gnn_exp, Li2021ivdetect} on explainable vulnerability detection are typically \textit{factual reasoning}-based explainers.
The core idea of these explainers is to identify key features in the input data (e.g., sub-graphs in the code graph) that contribute to the final predictions.
The selected features are commonly regarded as factual explanations,
as they derive from empirical input data and serve as factual evidence for particular outcomes.

Here, we contend that those factual reasoning-based explanations, which merely delineate the features or sub-graphs contributing to the identified vulnerability, are not convincing enough.
One reason is that developers remain uncertain about the actual influence of the code segments, which constitute the explanation sub-graph, on the detection result. 
In other words, the factual reasoning-based explanations cannot answer ``\textit{What would happen to the detection system's decision if we were to alter these code segments into alternative structures?}'' 
This perspective of \emph{what-if} is often associated with a human cognitive activity that imagines other possible scenarios for events that have already happened~\cite{roese1997cf_thinking}.
This motivates us to develop
a novel paradigm for analyzing detected vulnerabilities in source code - \emph{what-if} analysis.
In our cases, \emph{what-if} analysis explores hypothetical code instances with alternative structures. 
This approach aims to identify potential changes that would  fix the vulnerability, thereby providing a better explanation of the root causes and factors contributing to its existence.

\noindentparagraph{\textup{\textbf{Why \textit{What-If} Analysis? A Motivating Example.}}}
We use Figure~\ref{fig_illustration} as an example to illustrate the advantage of analyzing \emph{what-if} in explaining vulnerability detection compared to factual reasoning-based explanations.
This example involves a heap-based buffer overflow vulnerability in the \texttt{FFmpeg} project\footnote{\url{https://github.com/FFmpeg/FFmpeg}}, reported by CVE-2016-10190\footnote{\url{https://www.cvedetails.com/cve/CVE-2016-10190}}, which allows remote Web servers to execute arbitrary code via a negative chunk size in an HTTP response.
Specifically, this vulnerability arises from misuse of the \texttt{strtoll} function for parsing \texttt{chunksize} from HTTP responses into \texttt{int64\_t} format, without properly validating for negative values (\textcircled{4}).
Then, a negative \texttt{chunksize} can result in an erroneous calculation in the \texttt{FFMIN} function, producing a negative \texttt{size} for buffer operations (\textcircled{5}). 
This negative \texttt{size} potentially triggers out-of-bounds write operations, ultimately leading to a heap buffer overflow (\textcircled{7} and \textcircled{8}). 
In this example, the vulnerability detection system parses the code snippet into a semantic code graph (e.g., \textit{Abstract Syntax Tree} (AST), \textit{Control Flow Graph} (CFG), \textit{Data Flow Graph} (DFG), or \textit{Program Dependency Graph} (PDG)).
Here, without loss of generality, we consider the parsed code graph as a DFG for better illustration.

The vulnerability detection systems employ GNNs to model the DFG and yield a prediction outcome that classifies the input code snippet as vulnerable.
To explain the prediction of ``vulnerable'', the factual reasoning-based explanation identifies a compact sub-graph in the code graph (\textcircled{1}, \textcircled{5}, \textcircled{7}, and \textcircled{8}) as the key feature that contributes to the detected vulnerability.
This allows developers to recognize segments \textcircled{7} and \textcircled{8}, which involve buffer write operations (\textcircled{1} and \textcircled{5} are not involved), as potentially vulnerable blocks.
However, the explanation provided is inadequate for guiding code rectification to alter the detection system's decision, leaving developers to manually check variables such as \texttt{len}, \texttt{size}, and \texttt{chunksize} to identify the actual cause of the vulnerability.

In contrast, to investigate the context of vulnerability occurrences, \emph{what-if} analysis proactively and iteratively explores diverse hypothetical code structures (e.g., (a), (b), (c), and (d)), by inputting each into the detection system to observe varied prediction outcomes.
Taking structure (a) as an example, it is evident that removing the data-flow dependencies \textcircled{5}$\to$\textcircled{7} and \textcircled{5}$\to$\textcircled{8}, while retaining \textcircled{6}$\to$\textcircled{7}, leads to a prediction of ``non-vulnerable''.
It suggests that calculating \texttt{size} at \textcircled{5} may be a vulnerability source, while the computation of \texttt{len} at \textcircled{6} does not contribute to the vulnerability.
Through iterative exploration for subsequent structures (b), (c), and (d), \emph{what-if} analysis functions as an ``optimization'' process, eventually ``converging'' to a minimal change that alters the detection system's decision, i.e., only removing \textcircled{4}$\to$\textcircled{5} in structure (d).
The minimal change highlights the data flow \textcircled{4}$\to$\textcircled{5} as the root cause, which passes potentially incorrect \texttt{chunksize}, resulting in a miscalculation of \texttt{size} at \textcircled{5} and in turn triggering the buffer overflow at \textcircled{7} and \textcircled{8}. 
Consequently, developers receive an actionable insight, i.e., directly inspecting the value of \texttt{chunksize} at \textcircled{4} for potential errors.
Overall, we can conclude that the \emph{what-if} analysis essentially simulates the interactions between developers and the vulnerability detection system during debugging, methodically identifying the root causes of the detected vulnerabilities and guiding developers to effective solutions.

\vspace{-0.5em}
\noindentparagraph{\textup{\textbf{Our Solution and Contributions.}}}
Recent advances of \textit{counterfactual} reasoning in artificial intelligence~\cite{Abrate2021cf_graph_brain, Li2021causal_optimal_transport, Li2023causal_debias_recommendation, Lucic2022cfgnnexplainer, Tan2022cf2, Tan2021cf_rec, Wang2019mutli_agent_counterfactual, Yu2023deconfounded_recommendation} shed light on the possibility of applying \emph{what-if} analysis for GNN-based vulnerability detection.
A counterfactual instance represents an instance that, while closely similar to the original instance, is classified by the black-box model in a different class.
Thus, counterfactual reasoning aims to identify minimal changes in input features that can alter outcomes, thereby addressing the \emph{what-if} questions.

Building upon this motivation, we propose \systemnospace, the first explainer to introduce counterfactual reasoning for enhancing the explainability of GNNs in vulnerability detection.
Given a code instance, \system aims to identify a minimal perturbation to the code graph input that can flip the detection system's prediction from ``vulnerable'' to ``non-vulnerable''.
\system formulates the search problem for counterfactual perturbations as an edge mask learning task, which learns a differentiable edge mask to represent the perturbation.
Based on the differentiable edge mask, \system builds a counterfactual reasoning framework to generate insightful counterfactual explanations for the detection results.
Extensive experiments on four representative GNNs for vulnerability detection (i.e., GCN, GGNN, GIN, and GraphConv) validate the effectiveness of our proposed \systemnospace, both in terms of vulnerability-oriented and model-oriented metrics.

The key contributions of this paper are as follows.
\begin{itemize}[leftmargin=4mm, itemsep=0.05mm]
\item To the best of our knowledge, we are the first to discuss the \emph{what-if} question and introduce the perspective of counterfactual reasoning for GNN-based vulnerability detection.
\item We propose a counterfactual reasoning-based explainer, named \systemnospace, to generate explanations for the decisions made by the GNN-based vulnerability detection systems, which can help developers discover the vulnerability causes.
\item We conduct extensive experiments on four GNN-based vulnerability detection systems to validate the effectiveness of \systemnospace. 
Our results demonstrate that \system outperforms the state-of-the-art factual reasoning-based explainers.
\end{itemize}

\section{Background}

In this section, we begin by introducing essential preliminary knowledge necessary for a better understanding of our model. Subsequently, we present a mathematical formulation of the problem under study in this paper.

\subsection{GNN-Based Vulnerability Detection Model}
\label{sec_gnn_vul_detection}
Suppose that we have a set of $N$ code snippets 
$\mathcal{D} = \{C_1, C_2, \ldots, C_N\}$, 
and each code snippet $C_k$ is associated with a ground-truth label $Y_k \in \{0, 1\}$, which categorizes the code snippet as either non-vulnerable (0) or vulnerable (1).
The goal of vulnerability detection is to learn a mapping function $f(\cdot)$ that assigns a code snippet to either a non-vulnerable or vulnerable label. 

Current deep learning-based approaches follow a fundamental pipeline wherein the semantics of the source code are embedded into a hidden vector, which is then fed into a classifier.
Recently, GNNs have been designed to capture the semantic structures of source code, e.g., ASTs, CFGs, DFGs, and PDGs.
Given a code graph $G_k$ of $C_k$, GNN typically follows a two-step message-passing scheme (i.e., aggregate and update) at each layer $l$ to learn node representations for $G_k$.

Firstly, GNN generates an intermediate representation $\mathbf{m}_i^{l}$ for each node $i$ in $G_k$ by aggregating information from its neighbor nodes, denoted by $\mathcal{N}(i)$, using an aggregation function: 
\begin{equation}
    \mathbf{m}_i^{l} = \text{Aggregation}(\{\mathbf{h}_j^{l - 1} \mid j \in \mathcal{N}(i)\})\,,
\end{equation}
where $\mathbf{h}_j^{l - 1}$ denotes the representation of node $j$ in the previous layer. 
Subsequently, the GNN updates the intermediate representation $\mathbf{m}_i^{l}$ for each node $i$ via an update function: 
\begin{equation}
    \mathbf{h}_i^{l} = \text{Update}(\mathbf{m}_i^{l}, \mathbf{h}_i^{l - 1})\,.
\end{equation}
For a $L$-layer GNN, the final representation of the node $i$ is $\mathbf{h}_i^{L}$.
To obtain a graph representation $\mathbf{h}_k$ for the code graph $G_k$, a readout function (e.g., graph mean pooling) is applied to integrate all the node representations of $G_k$:
\begin{equation}
    \mathbf{h}_k = \text{Readout}(\{\mathbf{h}_i^{L}\})\,.
\end{equation}
Finally, the graph representation $\mathbf{h}_k$ is fed into a classifier (e.g. \textit{Multi-Layer Perception} (MLP)) followed by a Softmax function to calculate the probability distribution of non-vulnerable and vulnerable classes, as follows:
\begin{equation}
    P (c \mid G_k) = \text{Softmax}(\text{MLP}(\mathbf{h}_k))\,,
\end{equation}
where $P (c \mid G_k)$ is the predicted probability of the code snippet $C_k$ that belongs to each class in $\{0, 1\}$, i.e., $C_k$ is vulnerable or not.
The GNN model can be optimized by minimizing the binary cross-entropy loss between the predicted probabilities and the ground-truth labels, allowing it to learn from both non-vulnerable and vulnerable code instances in the training set.

As the model trained, in the testing phase, when presented with a code snippet $C_K$ accompanied by its code graph $G_k$, the trained GNN model $f(\cdot)$ is employed to compute the predicted probability $P(c \mid G_k)$ for each class. The resulting estimated label $\hat{Y}_k$ for $C_k$ is determined by selecting the class with the highest probability:
\begin{equation}
\hat{Y}_k = \argmax_{c \in \{0, 1\}} P (c \mid G_k)\,.
\label{equ_prediction_label}
\end{equation}

\noindentparagraph{\textbf{\textup{Investigated GNNs for Vulnerability Detection.}}}
In this study, we investigate four widely used GNNs for vulnerability detection.
These GNNs employ various implementations of the $\text{Aggregation}(\cdot)$ and $\text{Update}(\cdot)$ functions to capture structural code information for vulnerability detection.

\noindent$\triangleright$ \textbf{\textit{Graph Convolutional Network} (GCN)~\cite{KipfW2017GCN}} generalizes the idea of convolutional neural networks to graphs.
It aggregates neighbor node representations by summing them and utilizes an MLP to update the aggregated node representations.

\noindent$\triangleright$ \textbf{\textit{Gated Graph Neural Network} (GGNN)~\cite{Li2016ggnn}} utilizes a Gated Recurrent Unit~\cite{Cho2014rnn} to control information flow through edges when updating the aggregated node representations.

\noindent$\triangleright$ \textbf{\textit{Graph Isomorphism Network} (GIN)~\cite{Xu2019gin}} introduces the concept of graph isomorphism to ensure permutation invariance.
It employs a graph isomorphism operator to update the aggregated node representations.

\noindent$\triangleright$ \textbf{GraphConv~\cite{Morris2019graph_conv}} incorporates higher-order graph structures at multiple scales to enhance GNN's expressive power.

\begin{figure*}[!t]
    \centering
    \includegraphics[width=\textwidth]{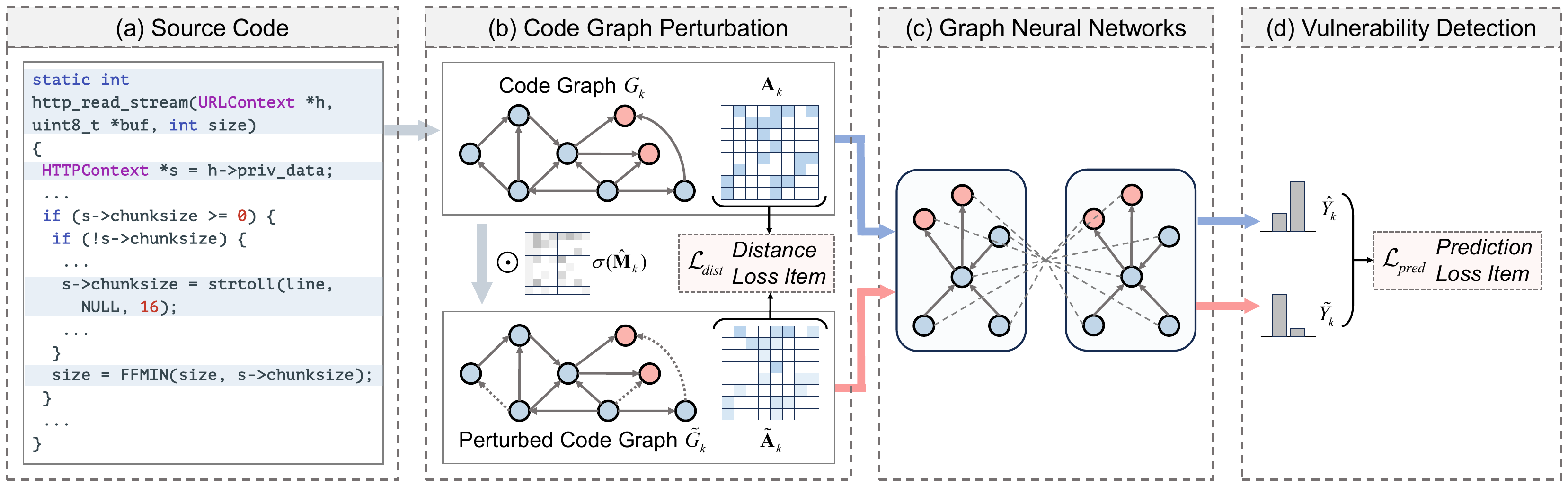}
    \vspace{-1.6em}
    \caption{
    An overview of our proposed counterfactual reasoning framework.
    }
    \label{fig_framework}
    \vspace{-0.4em}
\end{figure*}

\subsection{Model Explainability: The Problem}
Suppose that we have a trained GNN model $f(\cdot)$ and its prediction $\hat{Y}_k$ on the target code $C_k$ represented by a code graph $G_k$. 
In this paper, we explore the explainability of GNNs within a \textit{black-box} setting, recognized as a more challenging context for exploring model interpretability, where access to model parameters, training data, and gradients of each layer is unavailable.
In the black-box setting, we constrain the explainer to derive the prediction probability $P(c \mid G_k)$ exclusively by querying the model $f(\cdot)$ with the code graph $G_k$ as the input.

Under the aforementioned scenario, the factual reasoning-based explainers provide explainability by identifying key features that contribute to the model's prediction. 
For example, \citet{Li2021ivdetect} propose to seek a compact sub-graph $G_k^S$ that maintains the same prediction result as using the whole code graph $G_k$. They optimize the explainer by maximizing the probability of predicting the original estimated label $\hat{Y}_k$ when the input graph is limited to the sub-graph $G_k^S$, defined as:
\begin{equation}
\max_{G_k^S} P(\hat{Y}_k \mid G_k^S)\,.
\label{equ_conditional_entropy}
\end{equation}
On the contrary, counterfactual reasoning provides explainability by generating counterfactual instances to address \emph{what-if} questions. 
For the given code graph $G_k$, we generate its counterfactual instance by introducing a subtle perturbation to it, resulting in a new graph $\tilde{G}_k$. 
The perturbed graph $\tilde{G}_k$ differs minimally from the original $G_k$ but is classified in a different class, i.e., $f(\tilde{G}_k) \ne f(G_k)$.
As a result, counterfactual reasoning aims to identify a minimal perturbation to $G_k$ that alters the decision of the detection system.
We mathematically formulate the counterfactual reasoning problem as follows:
\begin{equation}
    \begin{gathered}
        \min_{\tilde{G}_k} d(\tilde{G}_k, G_k) \vspace{-1ex}\,, \\
        \text{s.t.},~\argmax_{c \in \{0, 1\}} P (c \mid \tilde{G}_k) \ne \hat{Y}_k\,,
    \end{gathered}
\label{equ_cf}
\end{equation}
where $d(\cdot, \cdot)$ represents a distance metric that quantifies the differences between $\tilde{G}_k$ and $G_k$, e.g., the number of edges removed by the perturbation.

\section{Proposed \system}

In this section, we propose a counterfactual reasoning-based explainer, named \systemnospace, for GNN-based vulnerability detection. 
\system comprises several key components: 
\textbf{(1) Code Graph Perturbation.}
\system employs a differentiable edge mask to represent the perturbation to the code graph, which transforms the discrete search task for counterfactual perturbations into a continuous learning task for edge masks. 
\textbf{(2) Counterfactual Reasoning Framework.}
Based on the differentiable edge mask, \system constructs a counterfactual reasoning framework and designs a differentiable loss function to make this framework optimizable, as illustrated in Figure~\ref{fig_framework}.
\textbf{(3) Counterfactual Explanation Generation.}
After optimization for the counterfactual reasoning framework, \system generates counterfactual explanations for the detection system's predictions.
We will elaborate on each component of \system in the following. 

\subsection{Code Graph Perturbation}
In our scenario, vulnerabilities often arise from incorrect or inconsistent structural relations in the source code, such as control and data flow flaws. Thus, for the given code graph $G_k$, we focus on perturbing its graph structures (i.e., edges), represented by the adjacency matrix $\mathbf{A}_k \in \{0, 1\}^{n \times n}$, rather than perturbing the node features $\mathbf{X}_k \in \mathbb{R}^{n \times d}$, where $n$ is the number of nodes in $G_k$ and $d$ represents the feature dimension. Note that the code graph $G_k$ is a directed graph, hence, $\mathbf{A}_k$ is unsymmetrical. 

One straightforward approach for generating counterfactual perturbations is through greedy search, which iteratively edits the code graph by removing or re-adding edges.
However, its practicality is limited by the vast size of the search space, leading to inefficiency~\cite{Abrate2021cf_graph_brain}.
Although heuristic strategies can potentially explore the search space more efficiently, identifying the optimal counterfactual instance with precision is challenging. 
Specifically, there is no guarantee that the counterfactual perturbation identified is the minimal one necessary.

\noindentparagraph{\textup{\textbf{Edge Mask-Based Perturbation.}}}
To overcome these limitations, inspired by prior work~\cite{Li2021ivdetect, Lucic2022cfgnnexplainer, Ying2019gnnexplainer}, we adopt the \textit{edge masking} technique.
This technique treats the searching for counterfactual perturbations as an edge mask learning task. 
The idea is that a perturbed graph $\tilde{G}_k$ can be derived by masking out edges from the original code graph $G_k$, as follows:
\begin{equation}
    \mathbf{\tilde{A}}_k = \mathbf{A}_k \odot \mathbf{M}_k\,,
\end{equation}
where $\mathbf{\tilde{A}}_k$ is the perturbed version of $\mathbf{A}_k$, $\mathbf{M}_k \in \{0, 1\}^{n \times n}$ is a binary edge mask matrix, and $\odot$ denotes element-wise multiplication. If an element $\mathbf{M}_{k,ij} = 0$, it indicates the edge $(i, j)$ is masked out in $\mathbf{A}_k$. 
As directly learning the binary edge mask matrix $\mathbf{M}_k$ is not differentiable, we relax $\mathbf{M}_k$ to continuous real values, which is $\mathbf{\hat{M}}_k \in \mathbb{R}^{n \times n}$. Then, as illustrated in Figure \ref{fig_framework}(b), the perturbed adjacency matrix is generated by:
\begin{equation}
    \mathbf{\tilde{A}}_k = \mathbf{A}_k \odot \sigma(\mathbf{\hat{M}}_k)\,,
\end{equation}
where $\sigma(\cdot)$ represents the sigmoid function that maps the edge mask into the range $[0, 1]$, allowing a smooth transition between the presence and absence of edges. 
As a result, starting from a randomly initialized edge mask matrix, $\mathbf{\hat{M}}_k$ can be optimized via gradient descent.
This approach enables a quicker and more precise determination of the minimal counterfactual perturbation compared to search-based strategies.

\subsection{Counterfactual Reasoning Framework}
\label{sec_cf}
We build a counterfactual reasoning framework to generate explanations for the predictions made by the GNN-based vulnerability detection system. The core idea of our proposed framework is to identify a minimal perturbation to the code graph that flips the detection system's prediction.
This is achieved by addressing a counterfactual optimization problem, which will be formulated in the following.

Suppose that we have a trained GNN model (whose weight parameter $\mathbf{W}$ is fixed and inaccessible) and the code graph $G_k$ for the target code snippet $C_k$.
We first apply the edge mask $\mathbf{\hat{M}}_k$ on the code graph $G_k$ to generate a perturbed graph, i.e., $\tilde{G}_k$.
Subsequently, as shown in Figure \ref{fig_framework}, we feed the original and perturbed code graphs into the GNN model to produce respective estimated labels:
\begin{equation}
    \begin{gathered}
        \hat{Y}_k = \text{GNN}(\mathbf{A}_k, \mathbf{X}_k \mid \mathbf{W}) \,, \\
        \tilde{Y}_k = \text{GNN}(\mathbf{\tilde{A}}_k, \mathbf{X}_k \mid \mathbf{W}) \,.
    \end{gathered}
\end{equation}
where $\mathbf{X}_k$ denotes the features of the nodes in $G_k$.
To identify a minimal counterfactual perturbation, we learn the edge mask $\mathbf{\hat{M}}_k$ based on the optimization objective of the counterfactual reasoning problem.
Specifically, we reformulate Eq.~(\ref{equ_cf}) as follows:
\begin{equation}
    \begin{gathered}
        \min_{\mathbf{\hat{M}}_k} d(\mathbf{\tilde{A}}_k, \mathbf{A}_k) \vspace{-1ex}\,,~
        \text{s.t.},~\tilde{Y}_k \ne \hat{Y}_k\,.
    \end{gathered}
\label{equ_cf2}
\end{equation}
Here, the constraint part aims to ensure that the new prediction $\tilde{Y}_k$ is different from the original prediction $\hat{Y}_k$, while the objective part aims to encourage that the perturbed adjacency matrix $\mathbf{\tilde{A}}_k$ is as close as possible to the original adjacency matrix $\mathbf{A}_k$.

Direct optimization of Eq.~(\ref{equ_cf2}) is challenging since both its objective and constraint parts are non-differentiable.
To address this, we design two differentiable loss function items to make the two parts optimizable, respectively.

\noindentparagraph{\textup{\textbf{Prediction Loss Item.}}}
To satisfy the constraint condition in Eq.~(\ref{equ_cf2}), we design a prediction loss item $\mathcal{L}_{pred}$ to encourage the detection system towards producing a different prediction when the original code graph $G_k$ is perturbed into $\tilde{G}_k$, as follows:
\begin{equation}
\mathcal{L}_{pred} = P(\hat{Y}_k \mid \mathbf{\tilde{A}}_k, \mathbf{X}_k) \,.
\label{equ_pred_loss}
\end{equation}
This loss item aims to minimize the likelihood that the perturbed graph $\tilde{G}_k$ will maintain the original prediction $\hat{Y}_k$, thereby maximizing the chances of achieving an altered prediction outcome.

\noindentparagraph{\textup{\textbf{Distance Loss Item.}}}
To address the objective part in Eq.~(\ref{equ_cf2}), we utilize binary cross entropy as a differentiable distance function to quantify the divergence between the original and perturbed adjacency matrixes, which is formulated as follows:
\begin{equation}
\mathcal{L}_{dist} = \text{BinaryCrossEntropy}(\mathbf{\tilde{A}}_k, \mathbf{A}_k) \,.
\label{equ_dist_loss}
\end{equation}
This distance function is chosen for its efficacy in measuring the difference between two probability distributions.
In our case, we consider the presence and absence of edges in the graph as binary classes, thus conceptualizing $\mathbf{\tilde{A}}_k$ as the estimated distribution of edges and $\mathbf{A}_k$ as the actual distribution.
During optimization, $\mathcal{L}_{dist}$ ensures that $\mathbf{\tilde{A}}_k$ remains as close as possible to $\mathbf{A}_k$, thus determining a minimal counterfactual perturbation to the code graph $G_k$.

\noindentparagraph{\textup{\textbf{Overall Loss Function.}}}
We integrate the above two loss items into an overall loss function to optimize them collaboratively:
\begin{equation}
\mathcal{L} = \alpha \cdot \mathcal{L}_{pred} + (1 - \alpha) \cdot \mathcal{L}_{dist} \,,
\label{equ_cf_loss1}
\end{equation}
where $\alpha$ is a hyper-parameter that regulates the trade-off between the prediction loss item and the distance loss item. 
Higher $\alpha$ prioritizes changing the prediction outcome, potentially at the expense of a larger perturbation, whereas lower $\alpha$ focuses more on minimizing the perturbation.
Based on the overall loss function, we optimize the counterfactual reasoning framework using the gradient descent algorithm and the Adam optimizer~\cite{Kingma2015adam}. 
Note that our framework operates in the \textit{black-box} setting, indicating that the process of counterfactual reasoning focuses solely on updating the edge mask $\mathbf{\hat{M}}_k$ to find the optimal perturbation while holding the underlying GNN model's parameters fixed.

\subsection{Counterfactual Explanation Generation}
Utilizing an optimized counterfactual reasoning framework, we generate counterfactual explanations to explain the predictions made by the vulnerability detection systems.

\noindentparagraph{\textup{\textbf{Generating Optimal Counterfatual Explanation.}}}
After optimization, we obtain the optimal edge mask $\mathbf{\hat{M}}_k^*$.
In this mask matrix, higher values indicate their corresponding edges should be preserved while lower values indicate their corresponding edges should be removed to reverse the detection system's decision. 
To form the final explanation, we employ a hyper-parameter $K_M$ to control the number of edges to be perturbed, i.e., taking the $K_M$ edges with the lowest mask values.
Then, we obtain the optimal counterfactual perturbed graph $\tilde{G}_k^{*}$ by removing the $K_M$ selected edges and derive a sub-graph:
\begin{equation}
    G_k^{S*} = G_k - \tilde{G}_k^{*}\,.
\end{equation}
As a result, the optimal counterfactual explanation takes the following form: 
the derived sub-graph $G_k^{S*}$ is the most critical factor on the detection result, that \textit{if removed, then the code would not be predicted as vulnerable}.

\noindentparagraph{\textup{\textbf{Deriving Diverse Counterfactual Explanations.}}}
In real-world scenarios, developers may need diverse counterfactual explanations to explore and understand the context of the detected vulnerability.
To achieve this, we build a narrowed search space based on the sub-graph $G_k^{S*}$.
Within this space, we employ exhaustive search to methodically explore and filter various edge combinations in $G_k^{S*}$ whose removal would alter the detection system's prediction. 
This process generates a set of diverse counterfactual explanations, each offering insights into the detected vulnerability from different perspectives.
Moreover, such diversity provides developers with multiple actionable options to address the detected vulnerability.

\section{Experimental Setup}
In this section, we begin by presenting the dataset, the baseline explainers for comparison, and the implementation details. Subsequently, we introduce two types of evaluation metrics to quantitatively evaluate the effectiveness of our proposed \systemnospace.

\subsection{Dataset}

Aligning with previous studies~\cite{Fu2022linevul, Hin2022linevd, Hu2023vul_interpreters, Li2021ivdetect}, we conduct our experiments on the widely-used vulnerability dataset, Big-Vul~\cite{Fan2020bigvul}.
Linked to the public CVE database~\cite{Gu2022hierarchical}, Big-Vul comprises extensive source code vulnerabilities extracted from 348 open-source C/C++ GitHub projects, spanning from 2002 to 2019.
It encompasses a total of 188,636 C/C++ functions, including 10,900 vulnerable ones, covering 91 various vulnerability types. 
Unlike other existing vulnerability datasets (i.e., Devign~\cite{Zhou2019devign} and Reveal~\cite{Chakraborty2022reveal}) which only provide vulnerability labels at the function level, Big-Vul offers more detailed, statement-level code changes derived from original git commits. These code changes for fixing vulnerabilities are crucial in our study. They enable us to build ground-truth labels for quantitatively evaluating the quality of the generated explanations (see Section~\ref{sec_exp_eval}).

To enhance the dataset's quality, we follow the cleaning procedure proposed by~\citet{Hin2022linevd}.
Specifically, we remove comment lines from the code and ignore purely cosmetic code changes (e.g., changes to whitespace).
We also exclude improperly truncated or unparsable code snippets. 
Additionally, following the practices of previous research~\cite{Fu2022linevul, Hin2022linevd}, we perform random undersampling for non-vulnerable code snippets to obtain a balanced dataset. 
In this work, we employ an open-source code analysis tool, Joern~\cite{joern, Yamaguchi14joern}, to parse each code snippet into a PDG, which serves as the input for the GNN-based detection model.
PDG is a commonly used graph representation for code in vulnerability detection research~\cite{Fu2022linevul, Hin2022linevd, Hu2023vul_interpreters, Li2021ivdetect}, which takes code statements as nodes and control-flow or data-flow dependencies as edges.
Finally, the dataset is randomly divided into training, validation, and testing sets with a ratio of 8:1:1.
Note that the explainers only generate explanations for the detection model's predictions on the test set.

\subsection{Baselines}
To provide a comparative analysis, we investigate six prominent factual reasoning-based GNN explainers as our baselines:

\begin{itemize}[leftmargin=4mm, itemsep=0.05mm]
    \item \textbf{GNNExplainer}~\cite{Lucic2022cfgnnexplainer} seeks a crucial sub-graph by maximizing the mutual information between the original GNN's prediction and the sub-graph distribution. 
    \item \textbf{PGExplainer}~\cite{Luo2020pgexplainer} learns an edge mask predictor based on the mutual information loss used in~\cite{Lucic2022cfgnnexplainer}. It accesses the training set to train the edge mask predictor.
    \item  \textbf{SubgraphX}~\cite{Yuan2021subgraphx} employs the Monte Carlo tree search algorithm~\cite{Silver2017monte} to efficiently identify important sub-graphs with a node pruning strategy.
    \item \textbf{GNN-LRP}~\cite{Schnake2022gnnlrp} decomposes the GNN's prediction scores into the importance of various graph walks using a higher-order Taylor decomposition and returns a set of most important graph walks as an explanation.
    \item \textbf{DeepLIFT}~\cite{Shrikumar2017deeplift} is another decomposition-based explainer but originally designed for image classification. 
    A previous work~\cite{Yuan2023gnn_exp_survey} extends it to explain GNN models, denoted as \textbf{DeepLIFT-Graph}.
    \item \textbf{GradCam}~\cite{Selvaraju2017gradcam} is a popular gradient-based explainer for image classification. It backpropagates the prediction scores to compute the gradients, which are then used to approximate the input importance. The previous work~\cite{Yuan2023gnn_exp_survey} adapts it for explaining GNN models, denoted as \textbf{GradCam-Graph}.
\end{itemize}

For the hyper-parameters of these baseline explainers, we adopt the implementation provided by previous research~\cite{Hu2023vul_interpreters, Yuan2023gnn_exp_survey}. 
Note that PGExplainer, GNN-LRP, DeepLIFT-Graph, and GradCam-Graph do not operate in the black-box setting, as they require access to model parameters, training data, and gradient information of GNNs.

\subsection{Implementation Details}
Our implementation comprises two main components: training GNN-based vulnerability detection models and generating explanations for the detection model's predictions.

\subsubsection{GNN-Based Vulnerability Detection}
In our experiments, we reimplement four vulnerability detection models employing different GNN cores (i.e., GCN, GGNN, GIN, and GraphConv). 
Each detection model adopts a two-layer GNN architecture with a hidden dimension of 256, followed by graph mean pooling to derive graph-level representations.
The graph-level representations are then input to a two-layer MLP classifier for vulnerability detection. 
In the model, we utilize GraphCodeBERT's token embedding layer~\cite{guo2020graphcodebert} to initialize node features for the input code graph.
ReLU activation functions are used after each layer, except for the final one, to introduce non-linearity. 
Based on the binary cross-entropy loss, we train each detection model using the Adam optimizer~\cite{Kingma2015adam} for 50 epochs, with a learning rate of 0.005 and a batch size of 64.
As shown in Tabel \ref{tab:gnn_performance}, following prior research~\cite{Cheng2021deepwukong, Li2018vuldeepecker, Li2021ivdetect}, we evaluate the performance of the reimplemented detection models using Accuracy, Precision, Recall, and $F_1$ score.
The results show that all four detection models achieve an Accuracy over 70\%, a Precision over 55\%, a Recall over 40\%, and an $F_1$ score over 50\%. 
Among them, GCN excels in Precision, while GIN leads in Accuracy, Recall, and $F_1$ score.
Overall, these models exhibit similar performance with high Precision and relatively low Recall. 

\subsubsection{Explanation Generation}
For the implementation of our proposed \systemnospace, we train it using Adam to minimize the loss function described in Section~\ref{sec_cf} for 800 epochs at a learning rate of 0.05. 
Note that, for each code snippet sample, \system is trained individually to explain the detection model's prediction.
We set the hyper-parameter $K_M$ to 8 by default and use the same $K_M$ value to control the size of the explanation sub-graphs generated by the factual reasoning-based explainers for fair comparison.
In addition, in Section~\ref{sec_parameter_analysis}, we conduct a parameter analysis on the hyper-parameter $\alpha$, exploring values from 0.1 to 0.9 to understand its influence on \systemnospace's performance.
It should be noted that the explainers aim to provide explanations by identifying the critical factors that contribute to the detected vulnerability. Thus, it is meaningless to explain the non-vulnerable code snippets and unfair to explain the code snippets that are incorrectly detected as vulnerable. As a result, we only consider explaining vulnerable code snippets that are correctly detected.

\begin{table}[!t]
    \centering
    \caption{The performance of the reimplemented GNN-based vulnerability detection models.}
    \begin{tabular}{l|cccc}
        \toprule
        GNN Core & Acc (\%) & Pr (\%) & Re (\%) & $F_1$ (\%) \\
        \midrule
        GCN & 72.05 & 60.39 & 44.81 & 51.44 \\
        GGNN & 71.89 & 59.43 & 47.08 & 52.54 \\
        GIN & 72.16 & 58.71 & 53.08 & 55.75 \\
        GraphConv & 70.98 & 56.61 & 52.11 & 54.27 \\
        \bottomrule
    \end{tabular}
    \label{tab:gnn_performance}
\end{table}

\begin{table*}[!t]
\normalsize
\setlength\tabcolsep{2pt}
\begin{minipage}[t]{\textwidth}
\caption{Comparison for the vulnerability-oriented evaluation results of explainers.}
\centering
\begin{tabular}{l|cccc|cccc|cccc|cccc}
    \toprule
    \multirow{2}{*}{Explainer} & \multicolumn{4}{c|}{GCN} & \multicolumn{4}{c|}{GGNN} & \multicolumn{4}{c|}{GIN} & \multicolumn{4}{c}{GraphConv} \\ 
    \cmidrule(lr){2-17}
     & Acc (\%) & Pr (\%) & Re (\%) & $F_1$ (\%) & Acc (\%) & Pr (\%) & Re (\%) & $F_1$ (\%) & Acc (\%) & Pr (\%) & Re (\%) & $F_1$ (\%) & Acc (\%) & Pr (\%) & Re (\%) & $F_1$ (\%) \\
    \midrule
    GNNExplainer & \underline{59.06} & \underline{13.68} & \underline{41.26} & \underline{17.29} & \textbf{61.25} & 13.94 & \textbf{45.54} & \textbf{18.76} & 53.37 & 12.14 & 34.42 & 15.09 & \underline{53.12} & \textbf{12.81} & \underline{37.54} & \underline{16.31} \\
    PGExplainer & 42.39 & 11.70 & 26.41 & 13.71 & 53.98 & 13.78 & 38.12 & 17.31 & 44.79 & 11.20 & 30.08 & 13.93 & 46.25 & 12.42 & 31.98 & 15.17 \\
    SubGraphX & 43.12 & 12.44 & 27.29 & 13.77 & 41.52 & 12.53 & 27.60 & 14.48 & 36.81 & 11.29 & 23.14 & 12.59 & 42.50 & 12.64 & 26.60 & 14.09 \\
    GNN-LRP & 56.00 & 13.31 & 38.52 & 16.49 & \underline{59.86} & 13.32 & 44.19 & 17.83 & \underline{54.94} & \underline{14.20} & \underline{39.54} & \underline{17.54} & 48.74 & 12.51 & 34.85 & 15.52 \\
    DeepLIFT-Graph & 50.00 & 12.88 & 33.14 & 15.61 & 55.36 & \textbf{14.39} & 39.83 & 17.84 & 47.24 & 12.89 & 32.84 & 15.58 & 49.69 & 12.48 & 34.85 & 15.43 \\
    GradCam-Graph & 44.93 & 12.93 & 27.69 & 14.54 & 56.06 & 13.22 & 41.04 & 17.23 & 44.17 & 13.62 & 30.03 & 15.64 & 41.88 & 11.73 & 28.91 & 13.96 \\
    \textbf{\systemnospace} & \textbf{61.23} & \textbf{13.84} & \textbf{42.84} & \textbf{17.84} & \textbf{61.25} & \underline{14.13} & \underline{44.30} & \underline{18.48} & \textbf{60.12} & \textbf{14.36} & \textbf{42.29} & \textbf{18.03} & \textbf{53.75} & \underline{12.77} & \textbf{38.36} & \textbf{16.32} \\
    \bottomrule
\end{tabular}
\label{tab:explainer_effectiveness}
\leftline{Note: We highlight the best score in \textbf{bold} and the second best score in \underline{underlined} in each column.}
\end{minipage}
\end{table*}

\subsection{Evaluating the Explainability}
\label{sec_exp_eval}
In this section, we introduce two types of metrics to evaluate the quality of the generated explanations quantitatively.

\subsubsection{Vulnerability-Oriented Evaluation Metric}

Evaluating counterfactual explanations in code is challenging due to the difficulty in obtaining standardized ground truth. 
Previous research~\cite{Cito2022cf_code} has relied on manual labeling for evaluation, which is costly, not easily scalable, and lacks standardization.
Fortunately, the Big-Vul dataset mitigates this issue by providing detailed statement-level fixes within git commits, which accurately reflect the changes addressing vulnerabilities. 
We utilize these commits to construct standardized ground-truth labels for our generated counterfactual explanations. 

In the vulnerability-oriented evaluation, following methodologies established in vulnerability detection research~\cite{Fan2020bigvul, Hin2022linevd, Hu2023vul_interpreters, Li2021ivdetect}, we adopt the statements that are deleted or modified in the commit (marked with ``\texttt{-}'' signs) as ground-truth labels.
Specifically, we extract all the statements from the vulnerable version of the code to build a binary ground-truth vector, denoted as $S = [s_1, s_2, \ldots, s_r]$, where $s_i = 1$ indicates the $i$-th statement is deleted or modified in the fixed version, and $s_i = 0$ otherwise.
Correspondingly, we construct a binary explanation vector $\Delta = [\delta_0, \delta_1, \ldots, \delta_r]$, where non-zero values in $\Delta$ represent the corresponding statements included in the generated explanation sub-graph. 
The comparison of $\Delta$ with the ground-truth vector $S$ allows for a quantitative evaluation of how accurately the generated explanations identify critical statements associated with the vulnerability.

Consider a given set of $M$ vulnerable code snippets denoted as $\{C_1, C_2, \ldots, C_M\}$ for evaluation.
For each code snippet, an explanation is deemed correct if it encompasses the deleted or altered statements. Consequently, we compute the Accuracy score by determining the percentage of accurate explanations among all generated explanations.
Moreover, we calculate the Precision and Recall scores for each code snippet by comparing the explanation vector $\Delta$ and the ground-truth vector $S$:
\begin{equation}
\text{Precision} = \frac {\sum_{i=1}^r s_i \cdot \delta_i} {\sum_{i=1}^r \delta_i}\,, \ \ \text{Recall} = \frac {\sum_{i=1}^r s_i \cdot \delta_i} {\sum_{i=1}^r s_i}\,.
\end{equation}
In our scenario, Precision measures the proportion of statements in the explanation that are relevant and accurately pertain to the vulnerability. On the other hand, Recall measures the proportion of ground-truth statements that are accurately included in the explanation. 
Additionally, we compute $F_1$ as the harmonic mean of the two scores to evaluate the overall performance. The formula for F1 is given as follows:
\begin{equation}
    F_1 = \frac {2 \cdot \text{Precision} \cdot \text{Recall}} {\text{Precision} + \text{Recall}} \,.
\end{equation}
Finally, we calculate the average scores of Precision, Recall, and $F_1$ across all code snippets.

\subsubsection{Model-Oriented Evaluation Metric}

The vulnerability-oriented evaluation metrics primarily focus on assessing the consistency between the generated explanations and the root causes of the detected vulnerabilities. 
However, these metrics cannot quantify to what extent the generated explanations really influence the detection system's decisions.
Thus, inspired by previous research~\cite{Tan2021cf_rec, Tan2022cf2}, our model-oriented evaluation borrows insights from causal inference theory and introduces \textbf{\textit{Probability of Necessity} (PN)}~\cite{glymour2016causal} to fill this gap. 
Intuitively, for an explanation $E$ that is generated to explain prediction $P$, if $E$ does not happen then $P$ will not happen, we say $E$ is a \textit{necessary} explanation for supporting the prediction $P$. 
The core idea of PN is that: if we imagine a counterfactual world where the explanation sub-graph $G_k^S$ did \textbf{not} exist in the original code graph $G_k$, then whether the corresponding code snippet $C_k$ would \textbf{not} be detected as vulnerable? 
This is critical for understanding the causal impact of the explanations on the prediction outcomes.
Following this idea, we define PN as the proportion of the generated sub-graph explanations that are \textit{necessary} to influence the detection system's predictions, as follows:
\begin{equation}
        \text{PN} = \frac1{M}{\sum_{k}^M {pn}_k}\,, \ \ \text{where } {pn}_k = 
        \begin{cases}
        1, & {\text{if}~\hat{Y}_k^\prime \ne \hat{Y}_k}\,, \\
        0, & {\text{else}}\,,
        \end{cases} \\
\label{equ_pn}
\end{equation}
where $\hat{Y}_k^\prime = \mathop{\arg\max}_{c \in \{0, 1\}} P (c \mid G_k - G_k^S)$ represents the prediction result for the code snippet $C_k$ when the explanation sub-graph $G_k^S$ is removed from the original code graph $G_k$.
If removing $G_k^S$ changes the prediction $\hat{Y}_k$, the explanation is considered necessary.

\section{Experimental Results}
To evaluate the performance of our counterfactual reasoning approach, we address the following \textit{Research Questions} (RQs):
\begin{itemize}[leftmargin=4mm, itemsep=0.05mm]
    \item \textbf{RQ1: Vulnerability-Oriented Evaluation.} 
    How well does \system perform in comparison with state-of-the-art factual reasoning-based explainers in identifying the root causes of the detected vulnerabilities?
    \item \textbf{RQ2: Model-Oriented Evaluation.} 
    How well does \system perform in comparison with state-of-the-art factual reasoning-based explainers in generating explanations that really influence the detection model's decision?
    \item \textbf{RQ3: Influence of Hyper-parameter $\alpha$.} How do different settings of the trade-off hyper-parameter $\alpha$ impact the performance of \system?
\end{itemize}

\begin{figure*}[!t]
	\centering
	\includegraphics[width=\textwidth]{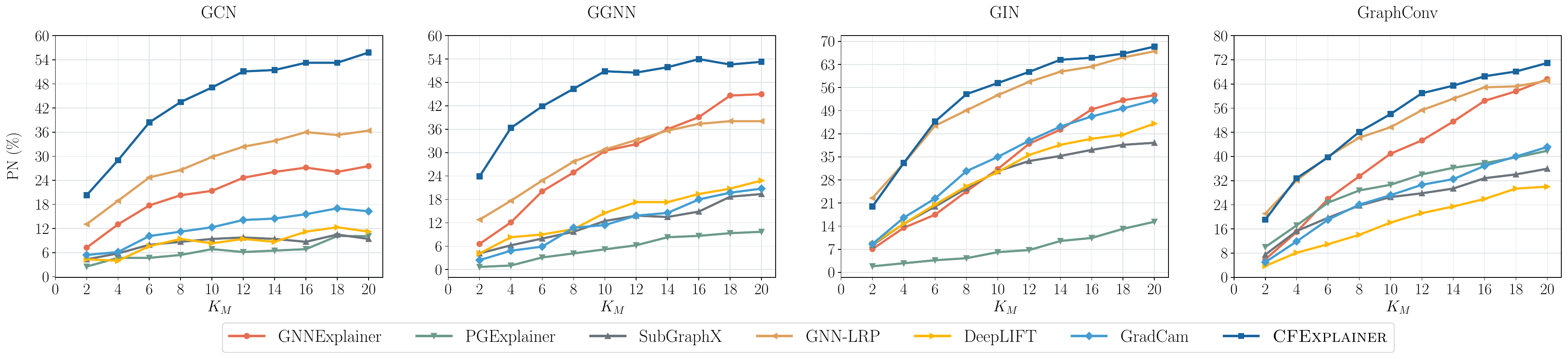}
	\caption{Comparison for the model-oriented evaluation results of explainers.}
	\label{fig_interpretability}
\end{figure*}

\subsection{RQ1: Vulnerability-Oriented Evaluation}

One of the key objectives of explainers in our context is to accurately identify the root causes of detected vulnerabilities. 
The effectiveness of our proposed \systemnospace, in comparison to factual reasoning-based explainers, is quantitatively showcased in Table \ref{tab:explainer_effectiveness}, which reports the vulnerability-oriented evaluation results on four GNN-based detection models: GCN, GGNN, GIN, and GraphConv.
These results reveal that \system outperforms the baseline explainers in most scenarios, demonstrating the effectiveness of our counterfactual reasoning approach. Across the four GNN-based detection models, \system achieves average improvements of 
24.32\%, 12.03\%, 28.22\%, and 14.29\% in Accuracy,
7.93\%, 4.43\%, 14.36\%, and 2.72\% in Precision,
32.28\%, 12.47\%, 33.51\%, and 18.19\% in Recall,
17.10\%, 7.18\%, 19.71\%, and 8.22\% in $F_1$ score
over the factual reasoning-based explainers.

Among all baseline explainers, the perturbation-based GNNExplainer exhibits relatively good performance by directly searching for a crucial sub-graph that significantly contributes to the vulnerability detected by GNNs. 
Besides, the decomposition-based methods (i.e., GNN-LRP and DeepLIFT-Graph) directly decompose the detection model's predictions into the importance of edges in the code graph and select the most important edges as an explanation, resulting in slightly inferior performance compared to GNNExplainer.
However, the other two perturbation-based methods (i.e., PGExplainer and SubGraphX) and the gradient-based GradCam-Graph method perform relatively poorly. 
This is because PGExplainer's mask predictor may suffer from the distribution shift between the training and test sets, while SubGraphX's node pruning strategy may be not compatible with our scenario of perturbing edges in the code graph.
GradCam-Graph utilizes gradient values to measure the edge importance, leading to an explanation sub-graph that correlates with the detection model's hidden information rather than the actual vulnerabilities.
In contrast to these factual reasoning-based explainers, \system aims to address \emph{what-if} questions by seeking a minimal perturbation to the code graph that alters the detection model's prediction from ``vulnerable'' to ``non-vulnerable''.
Through this exploration, \system delves deeply into the context where the vulnerability occurs, revealing causal relationships between code structures and detection outcomes, thereby discovering the root causes of the detected vulnerabilities.

\begin{tcolorbox}
\textbf{Answer to RQ1:} \system exhibits superior effectiveness in vulnerability-oriented evaluation, outperforming state-of-the-art factual reasoning-based explainers. 
\end{tcolorbox}

\begin{figure}[!t]
	\centering
	\includegraphics[width=0.98\linewidth]{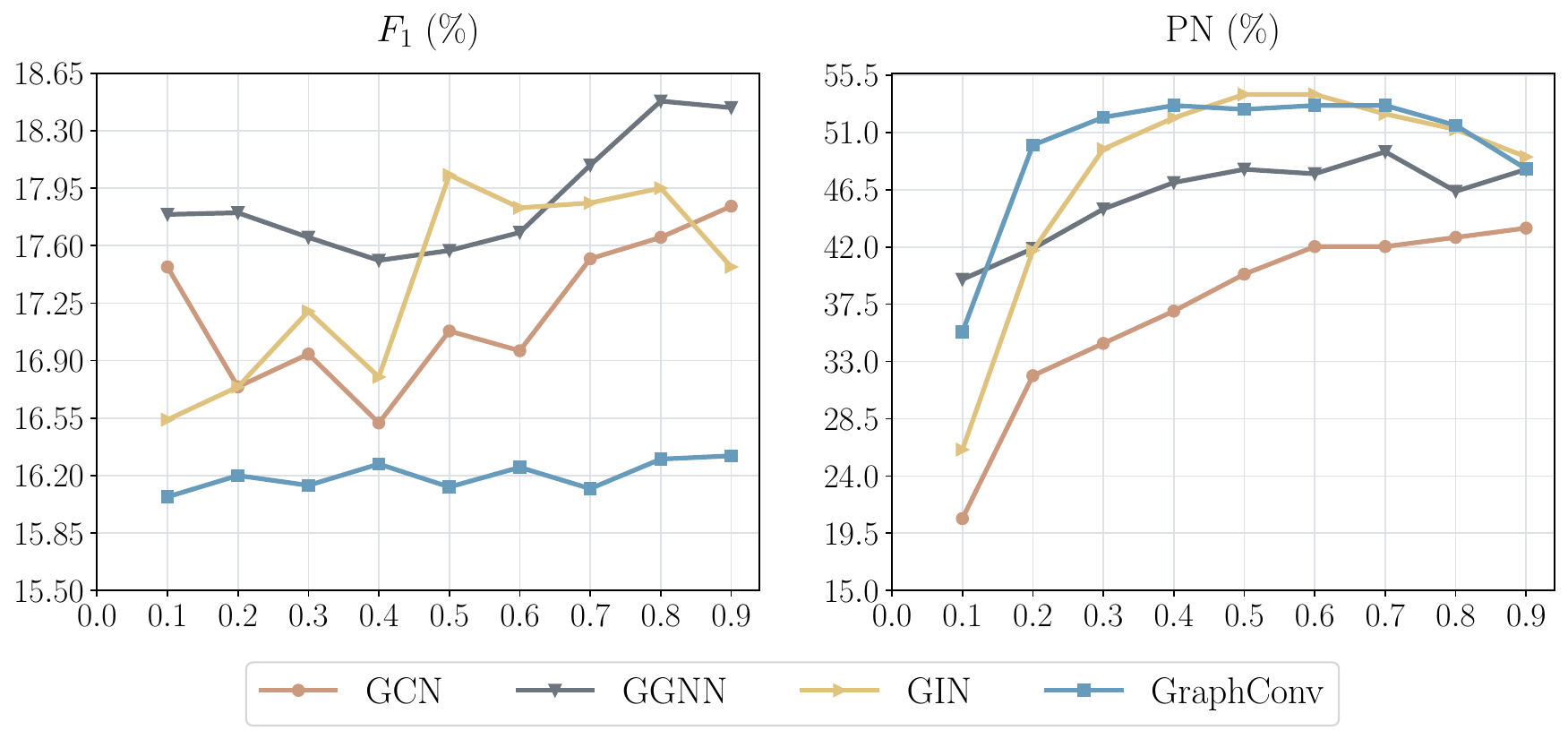}
	\caption{A parameter analysis on the hyper-parameter $\alpha$.}
	\label{fig_parameter_analysis}
\end{figure}

\subsection{RQ2: Model-Oriented Evaluation}
\label{sec_interpretability}

Compared to vulnerability-oriented evaluation, model-oriented evaluation focuses on assessing the \textit{necessity} of the generated explanations for supporting the detection model’s predictions. 
As illustrated in Figure \ref{fig_interpretability}, \system demonstrates superior performance over state-of-the-art factual reasoning-based explainers across four GNN-based detection models. Notably, the $\text{PN}$ curve for \system consistently encompasses those of the baseline explainers under various $K_M$ settings, visually indicating its effectiveness.
Unlike factual reasoning-based explainers that identify crucial sub-graphs but fall short in determining their actual influence on detection outcomes, \system targets a minimal change to the code graph that alters the prediction. 
This approach ensures the identification of edges that are truly necessary for the prediction outcome. 
This distinction is crucial in understanding \systemnospace's ability to provide more accurate and essential explanations.
Moreover, we can observe that as the value of $K_M$ increases, the $\text{PN}$ scores of all explainers generally show improvement.
This improvement can be attributed to that with a higher $K_M$ value, more crucial edges necessary for supporting the detection result are identified and included in the generated explanation sub-graph.

\begin{tcolorbox}
\textbf{Answer to RQ2:} \system demonstrates superior performance in model-oriented evaluation, outperforming state-of-the-art factual reasoning-based explainers. 
\end{tcolorbox}

\begin{figure*}[!t]
	\centering
	\includegraphics[width=\textwidth]{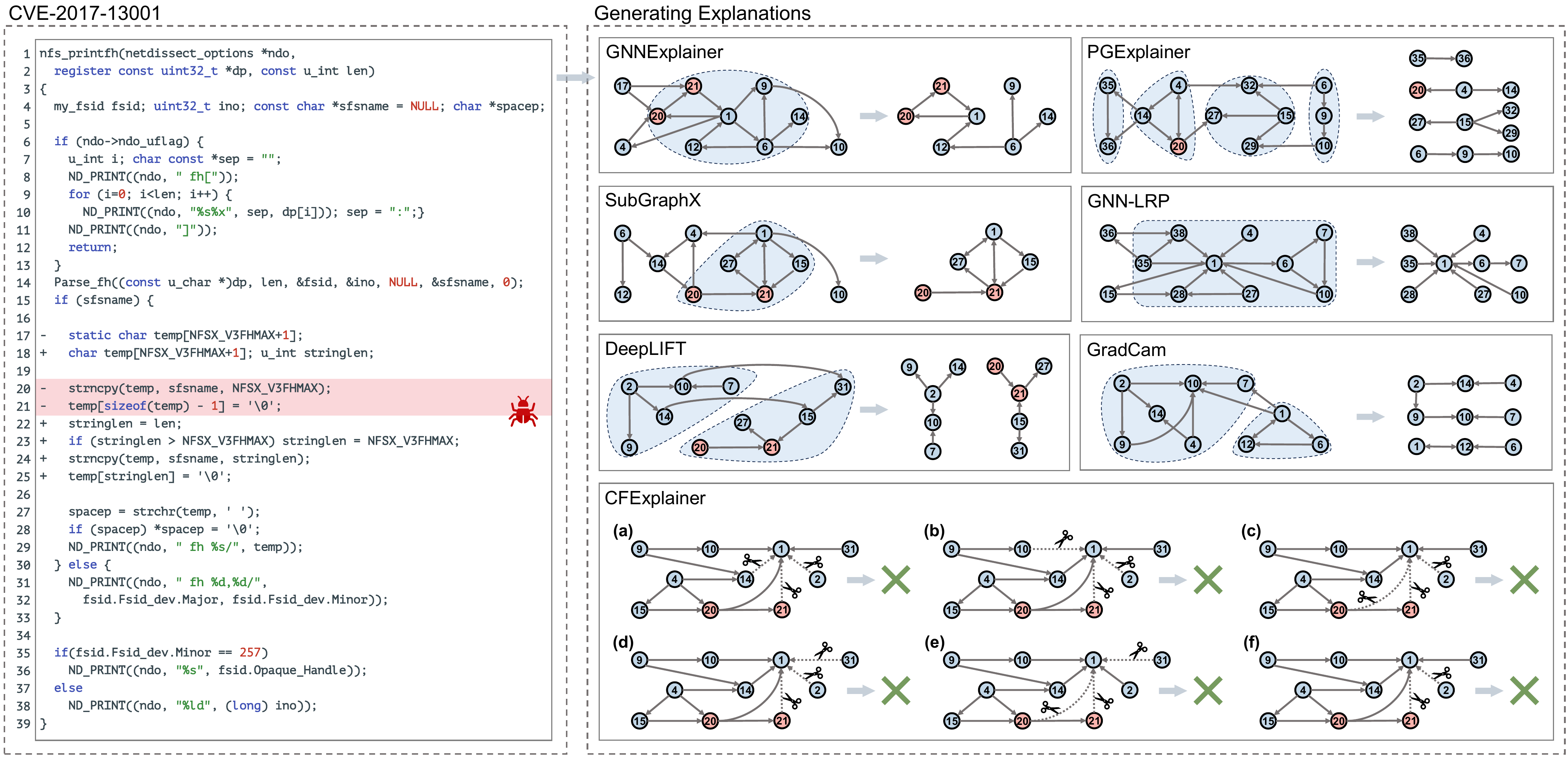}
	\caption{A case study on the CVE-2017-13001 vulnerability in the \texttt{tcpdump} project.}
	\label{fig_casestudy}
\end{figure*}

\subsection{RQ3: Influence of Hyper-parameter $\alpha$}
\label{sec_parameter_analysis}

Understanding the impact of the trade-off hyper-parameter $\alpha$ is crucial for optimizing \systemnospace's performance in generating counterfactual explanations.
The hyper-parameter $\alpha$ plays a pivotal role in balancing the emphasis between the prediction loss item and the distance loss item. 
We conduct the parameter analysis on \system by varying $\alpha$ from 0.1 to 0.9 while keeping other hyper-parameters fixed.

As shown in Figure~ \ref{fig_parameter_analysis}, $\alpha$ significantly influences the effectiveness of the counterfactual explanation generated by \systemnospace. 
We can see that with the increase in $\alpha$, the differences between the predictions using the perturbed graph and the original graph are more encouraged, thereby pushing the explanation to be more counterfactual, which leads to dramatic performance improvements. 
However, after $\alpha$ reaches its optimal value, the performance begins to decline because \system tends to generate a counterfactual larger perturbation to the code graph that may fail to identify the most critical factor influencing the detection system's prediction.
Based on our parameter analysis, we set $\alpha=0.9$ for GCN and GraphConv, $\alpha=0.8$ for GGNN, and $\alpha=0.5$ for GIN.
These values are chosen to ensure optimal performance across different models, accommodating their unique characteristics and sensitivities to the balance between prediction and distance loss.

\begin{tcolorbox}
\textbf{Answer to RQ3:} The trade-off hyper-parameter $\alpha$ has a significant impact on \systemnospace's performance. Optimal settings vary across models, with $\alpha=0.9$ for GCN and GraphConv, $\alpha=0.8$ for GGNN, and $\alpha=0.5$ for GIN.
\end{tcolorbox}

\subsection{Case Study}
We conduct a case study to qualitatively assess the effectiveness of \system compared to factual reasoning-based explainers, as shown in Figure~\ref{fig_casestudy}.
This case study involves a specific code commit of the \texttt{nfs\_printfh} function in the \texttt{print-nfs.c} file from the \texttt{tcpdump} project\footnote{\url{https://github.com/the-tcpdump-group/tcpdump}}. 
The added lines are indicated with a ``\texttt{+}'' sign, while the deleted lines are marked with a ``\texttt{-}'' sign. 
This commit addresses a buffer over-read vulnerability of the NFS parser, reported by CVE-2017-13001\footnote{\url{https://www.cvedetails.com/cve/CVE-2017-13001}}. 
This vulnerability arises from the original code failing to ensure that the source string \texttt{sfsname} is shorter than the destination buffer \texttt{temp} (\doublecircled{20}).
As a result, \texttt{strncpy} could copy more characters than \texttt{temp} can safely contain, without null-terminating it immediately after the last copied character (\doublecircled{21}). 
This leads to potential buffer over-reads when \texttt{temp} is later accessed as a string.
To address this vulnerability, the code should copy no more than \texttt{temp} can hold and must explicitly null-terminate the buffer after the last character copied from \texttt{sfsname}. 
The fix introduced in the commit ensures that the length of the data copied does not exceed \texttt{NFSX\_V3FHMAX} (\doublecircled{23}) and that \texttt{temp} is correctly null-terminated after the copy (\doublecircled{25}), preventing any over-read.

In this case study, we observe that factual reasoning-based explainers like GNN-LRP and GradCam fail to identify the vulnerability cause (\doublecircled{20} and \doublecircled{21}) in their generated explanation sub-graphs.
While other factual reasoning-based explainers effectively point out \doublecircled{20} or \doublecircled{21}, their generated explanations also contain a few code statements unrelated to the vulnerability.
This dilutes the clarity of the explanations, leaving developers to manually check the code to find out the actual vulnerability cause.
Conversely, \system excels by generating a set of diverse counterfactual explanations to help developers understand the context of the detected vulnerability.
For example, in structure (c), \system identifies that removing the program dependencies \textcircled{2}$\to$\textcircled{1}, \doublecircled{20}$\to$\textcircled{1}, and \doublecircled{21}$\to$\textcircled{1} alters the detection result.
This directly leads developers to the critical statements \doublecircled{20} and \doublecircled{21}, which involve buffer operations (\textcircled{1} and \textcircled{2} are not involved), and identifies them as potential areas of the vulnerability.
Further, through the analysis of the minimal counterfactual perturbation as in structure (f), \system offers an actionable insight, i.e., inspecting the null-terminating operation at \doublecircled{21} for potential errors.

\section{Threats to Validity}

The threats to the validity of our work are discussed as follows.

\noindentparagraph{\textup{\textbf{On the GNN Performance.}}}
The effectiveness of the counterfactual explainer is heavily influenced by the detection performance of the GNN model. Since our explainer generates explanations by perturbing the code graph instance, it relies on a reliable detection model to determine whether the perturbed instance is vulnerable or not. If the detection model has learned biased patterns and fails to produce the correct detection result for the perturbed instance, it can undermine the effectiveness of the explainer. 
Thus, to ensure the optimal performance of the explainer, we recommend using it in conjunction with GNN-based detection models that exhibit ideal detection performance. By combining the counterfactual explainer with high-performing detection models, we can enhance the overall effectiveness and reliability of the explanation process.

\noindentparagraph{\textup{\textbf{On the Perturbation of Code Graphs.}}}
Our current counterfactual explainer is mainly based on graph theory principles and does not specifically consider the unique features of vulnerabilities. In our future work, we plan to enhance our explainer by incorporating perturbation algorithms specifically tailored to the vulnerabilities in code graphs. 
This will enable us to achieve a more specialized counterfactual explainer, which can better capture the underlying characteristics of vulnerabilities and provide more accurate insights into the behavior of GNN-based vulnerability detection models.

\section{Related Work}
In this section, we review the related literature about vulnerability detection and localization, explainability in software engineering, and counterfactual reasoning in GNNs.

\subsection{Vulnerability Detection and Localization}
Vulnerability detection plays a crucial role in ensuring the security and reliability of software systems. 
Existing efforts in vulnerability detection can be generally divided into two main approaches: 
static analysis-based~\cite{Gao2018cobot, Sui2016svf, Viega2000its4} and deep learning-based~\cite{Chakraborty2022reveal, Cheng2021deepwukong, Li2018vuldeepecker, Zhou2019devign} approaches.
Traditional static analysis-based approaches require human experts to manually define specific rules, which suffers from efficiency issues.
On the other hand, deep learning-based approaches have gained increasing interest in vulnerability detection, due to their strong capability in representing the semantics of source code.
However, compared with static analysis-based approaches, the deep learning-based approaches cannot provide a fine-grained analysis of which lines of the code may cause the detected vulnerabilities.

Although fault localization techniques like spectrum-based methods~\cite{de2016spectrumfl, Keller2017spectrumfl} and delta debugging~\cite{Zeller2002isolating, Zeller2002simplifying} could be employed to locate vulnerable code statements, their effectiveness relies on either the availability of extensive test suites or numerous time-consuming testing executions.
Recently, several deep learning-based line-level detection methods~\cite{Ding2022velvet, Fu2022linevul, Hin2022linevd, Li2022vuldeelocator, Zou2022mvulpreter} have been proposed to predict which statements in the code are vulnerable. 
However, these methods not only require large training samples to train the deep-learning models but also lack explainability in why certain statements are predicted as vulnerable.
Consequently, explainable approaches have attracted increasing attention in vulnerability detection.
Existing explainable approaches are mainly based on factual reasoning, which aims to find the input features that play a crucial role in the detection model's prediction~\cite{Li2021ivdetect, Ganz2021vul_gnn_exp, Hu2023vul_interpreters, Zou2021vul_inter_search}.
However, these approaches are limited in their ability to provide further insights on how to alter the detection model's prediction, especially when the code is predicted as vulnerable.
In contrast to the previous work, \system introduces counterfactual reasoning to identify what input features to change would result in a different prediction, thereby providing actionable guidance for developers to address the detected vulnerabilities.

\vspace{-0.4em}
\subsection{Explainability in Software Engineering}
Explainability poses a challenging issue in software engineering, especially due to the increasing dependence of developers on using the predictions provided by deep learning models to optimize their codes.
Recently, many efforts have been made to improve the explainability of deep learning models in software engineering~\cite{Bui2019autofocus, Cito2021rule_exp, Cito2022cf_code, Rabin2021simplification, sharma2023interpreting, Suneja2021probing, Wan2022capture}.
For instance, \citet{Cito2021rule_exp} focused on global explainability, which aims to find specific input data types on which the model exhibits poor performance. 
\citet{sharma2023interpreting} introduced a neuron-level explainability technique to identify important neurons within the neural network and eliminate redundant ones. 
\citet{Wan2020attention_code_retrieval} addressed the structural information of source code under a multi-modal neural network equipped with an attention mechanism for better explainability.
\citet{Wan2022capture} investigated the explainability of pre-trained language models of code (e.g., CodeBERT and GraphCodeBERT), which conducts a structural analysis to explore what kind of information these models capture.
Furthermore, several factual reasoning approaches have been proposed recently. 
For example, AutoFocus~\cite{Bui2019autofocus} employed attention mechanisms to rate and visualize the importance of code elements.
\citet{Zou2021vul_inter_search} proposed an explainable approach based on heuristic searching, aiming to identify the code tokens contributing to the vulnerability detector's prediction.
In addition, two previous studies~\cite{Rabin2021simplification, Suneja2021probing} proposed model-agnostic explainers based on program simplification techniques, which aims to simplify the input code while preserving the model's prediction results, inspired by the delta debugging algorithms~\cite{Zeller2002isolating, Zeller2002simplifying}.
Counterfactual reasoning has also been explored by a recent work~\cite{Cito2022cf_code}, similar to our work. However, this work focused on perturbing the plain text input of code to generate counterfactual explanations, in contrast to our work focusing on perturbing the graph input of code.

\subsection{Counterfactual Reasoning in GNNs}
Recently, several studies have explored the use of counterfactual reasoning to provide explanations for GNNs~\cite{Abrate2021cf_graph_brain, Bajaj2021cf_graph_robust, Huang2023global_cf, lin2021gem, Lucic2022cfgnnexplainer, ma2022clear, numeroso2021meg, Tan2022cf2, Wang2024reinforced_cf_recommendation, Wang2022mgpolicy, Wellawatte2022cf_molecule, Yu2023counterfactual_conversational_recommendation, Yu2023causality_graph_recommendation}.
For example, \citet{Lucic2022cfgnnexplainer} generated counterfactual explanations by identifying a minimal perturbation to a node's neighborhood sub-graph that would change the GNN's prediction on this node.
\citet{lin2021gem} employed Granger causality for counterfactual reasoning to learn explanations based on an auto-encoder model via supervised learning.
\citet{Bajaj2021cf_graph_robust} identified robust edge subsets whose removal would alter the GNN's predictions by learning the implicit decision regions in the graph.
\citet{ma2022clear} utilized a graph variational autoencoder for the optimization and generalization of counterfactual reasoning on graphs.
\citet{Tan2022cf2} incorporated both counterfactual and factual reasoning perspectives from causal inference theory.
\citet{Huang2023global_cf} explored global counterfactual reasoning for GNNs' global explainability.
Furthermore, counterfactual reasoning has been applied in domain-specific graph scenarios, such as molecular graphs~\cite{numeroso2021meg, Wellawatte2022cf_molecule} and brain networks~\cite{Abrate2021cf_graph_brain}.
While the idea of counterfactual reasoning in these studies is similar to our work, we are the first to investigate counterfactual reasoning on code graphs and provide counterfactual explanations for the vulnerability detection task.

\section{Conclusion}
In this paper, we propose \systemnospace, a novel counterfactual reasoning-based explainer for explaining the predictions made by GNN-based vulnerability detection models. 
\system generates counterfactual explanations by identifying the minimal perturbation to the code graph that can alter the detection system's prediction, thus addressing \emph{what-if} questions for vulnerability detection.
The counterfactual explanations can identify the root causes of the detected vulnerabilities and provide actionable insights for developers to fix them. 
Our extensive experiments on four GNN-based vulnerability detection models show that \system outperforms the existing state-of-the-art factual reasoning-based explainers.

The application of counterfactual reasoning in software engineering, particularly in the domain of vulnerability detection, is still in its early stages, offering substantial opportunities for further exploration. 
The success of \system encourages us to explore its application in broader tasks, including but not limited to bug detection, code search, and code clone detection. 
We believe that the principles of counterfactual reasoning can be effectively adapted to these areas, potentially transforming the way developers interact with and understand software systems.

\noindentparagraph{\textup{\textbf{Data Availability.}}}
All the experimental data and code used in this paper are available at 
\texttt{\url{https://github.com/CGCL-codes/naturalcc/tree/main/examples/counterfactual-vulnerability-detection}}.

\section*{Acknowledgment}
This work is supported by the Major Program (JD) of Hubei Province (Grant No. 2023BAA024).
We would like to thank all the anonymous reviewers for their insightful comments.

\balance
\bibliographystyle{ACM-Reference-Format}
\bibliography{ref}

\end{document}